# Deep and fast free-space electro-absorption modulation in a mobility-independent graphene-loaded Bragg resonator

Spyros Doukas, Alma Chatzilari, Alva Dagkli, Andreas Papagiannopoulos, Elefterios Lidorikis*

Department of Materials Science and Engineering, University of Ioannina, 45110 Ioannina, Greece

*elidorik@cc.uoi.gr

Deep and fast electro-optic modulation is critical for high-speed near infrared signal processing. We combine the electro-absorption tunability of graphene with the high-Q resonance of a Bragg-based Fabry-Perot resonator at $\lambda$=1550 nm and show that ~100% free-space signal modulation (>50 dB extinction ratio, <1 dB insertion loss) at high speed (>1 GHz) can always be achieved independently of graphene quality (mobility), provided the device is operating in reflection mode and tuned in critical coupling with graphene. Remarkably, the critical coupling mechanism produces a higher extinction ratio for lower graphene mobility. We use practical considerations to optimize the device architecture and operation as a function of graphene mobility. With a small modification this scheme can be turned into a very sensitive acousto-absorption modulator with a ~30 dB/Å extinction ratio, or an index sensor with $10^7$ %/RIU sensitivity. These designs can be easily extended throughout the midIR spectrum by appropriate scaling of layer thicknesses.

Photonic components for fast signal modulation and detection at near infrared (NIR) frequencies are important for applications in optical signal processing and communications[1], in both integrated[2,3] and free-space operation[4]. In the latter case, the $\lambda = 1550$ nm wavelength also appears as an optimal choice[5]. Graphene[6], with its electrically tunable conductivity[7] and fast carrier relaxation times (~ps)[8] is poised to make important contributions in this field[9,10]. Graphene-based NIR optical modulation has been demonstrated in waveguide-integrated[11-15] and in free-space [16-19] configurations. In the latter case up to ~70% modulation depth (MD) at ~MHz modulation frequency (MF) was recently demonstrated[17,18].

Graphene absorption is highly tunable within the Vis-midIR range *via* electrostatic gating, being ~2.3%[20,21] (when air-suspended and 2.3/$n$% when embedded in a dielectric[22] of index $n$) for $\hbar\omega > 2E_F$ due to interband transitions ($E_F$ is the graphene Fermi level determined by the electrostatic doping) and significantly reduced below this limit due to Pauli-blocking[23,24]. For much smaller frequencies it picks up again due to intraband transitions giving rise to graphene plasmonics[25-27]. Transitioning across the Pauli blocking point is relatively sharp for high quality graphene but smoothes up as the electron relaxation time gets shorter. In any case however, because of the overall small graphene absorption (<2.3%), it is imperative that the graphene-light interaction is increased by use of a resonator structure, such as cavity[16-19,28-32] or plasmonics[15,22,33-38].

Numerical simulations predict that free-space modulators consisting of graphene inserted in a Bragg-type resonator can reach MDs ~100%, with low insertion loss (IL) and ultrafast (~GHz) MFs[39,40], promising great opportunities in free-space NIR (or midIR-THz[39]) modulation. In these studies, however, a particular hypothesis for graphene quality (i.e. charge carrier relaxation time $\tau$ and mobility $\mu$) is adopted and the device architecture is optimized accordingly. Besides being problematic for devices operating in transmission mode (as we will show) since low graphene mobility will drastically reduce MD and increase IL, graphene quality consistency issues without a systematic understanding and design strategy around them can potentially turn into show-stoppers. In this work we show that a graphene-loaded Bragg resonator device operating in reflection mode can always be dynamically tuned into critical coupling[41-44] (total absorption) irrespective of graphene quality, ensuring systematically large MD (~100%, >50 dB extinction ratio), low IL (<1 dB, <0.01 dB for high mobility graphene) and high MF (>1 GHz). We use practical considerations to optimize the device architecture and operation as a function of graphene mobility. We also show that with a small modification this scheme can be turned into a very sensitive acousto-absorption modulator with a ~30 dB/Å extinction ratio, or an index sensor with $10^7$ %/RIU sensitivity. All the considerations and designs studied here for $\lambda = 1550$ nm can also be easily extended throughout the midIR spectrum by appropriately choosing the materials and scaling their thicknesses.

We assume the asymmetrical Bragg cavity shown in Fig. 1a consisting of 3 periods of a Si/SiO$_2$ bilayer on either side with an Au mirror on the back. The layer thicknesses are $d_{Si} = 113.3$ nm and $d_{SiO_2} = 265.4$ nm corresponding to quarter-wave layers at $\lambda = 1550$ nm for refractive indices $n_{Si} = 3.42$ and $n_{SiO_2} = 1.46$ respectively. The cavity consists of a SiO$_2$ double layer with a single layer graphene (SLG) in between. We model the graphene permittivity by the Kubo formula[45] using electron relaxation time $\tau = 200$ fs and assume that it can be electrostatically doped by interrogating the bottom Si layer (assuming all other Si layers intrinsic), as depicted in Fig. 1a (our conclusions are apparently valid for any other choice of materials comprising the Bragg stacks). The Au mirror at the bottom ensures there is no transmitted wave.

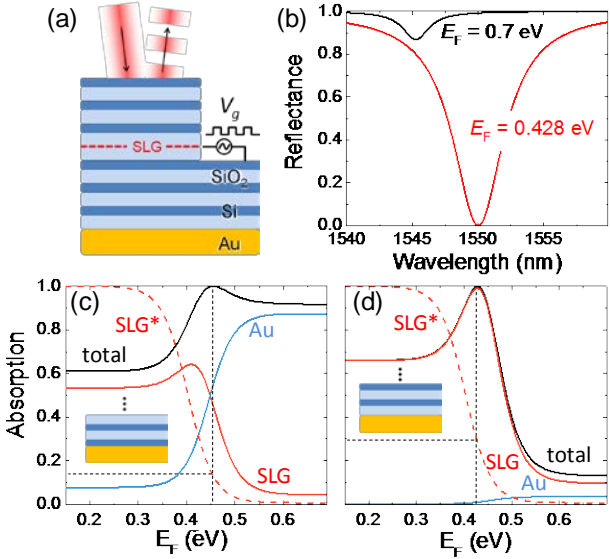

**Figure 1.** (a) Schematic of the proposed device with three Si/SiO$_2$ periods on either side of the cavity. (b) Reflectance of device in (a) for graphene electrostatic doping 0.4283 and 0.7 eV. (c-d) The absorption in each constituent as a function of graphene doping for a device having the Au mirror in contact with (c) Si and (d) SiO$_2$. The red dashed line marked as SLG* is the normalized absorptivity of suspended graphene.

The system reflection under normal incidence is calculated by employing the Fresnel relations with the transfer matrix method and plotted in Fig. 1b for two different SLG doping. There exists a critical doping level $E_F^c$ for which the reflection at $\lambda = 1550$ nm is practically zero. We note that in the absence of transmission this means perfect absorption, a condition also called critical coupling[41-44]. At any other doping level the reflection becomes finite and generally close to 1, yielding modulation depth ~100%. Our device is essentially a single-port system, given the negligible transmission imposed by the back mirror. In this case there is always a perfect absorption (critical coupling) condition as can be seen by inspection of the general single-port coupled-mode formulation for absorption:[44,46]

$$A = \frac{4\gamma_a\gamma_d}{(\omega - \omega_0)^2 + (\gamma_a + \gamma_d)^2}$$

where $\omega_0$ is the cavity resonance frequency, $\gamma_a$ the total absorption rate of the system and $\gamma_d$ the decay rate of the cavity. In our example absorption is provided by the grapheme layer in the cavity center and by the Au mirror in the back. At resonance ($\omega = \omega_0$) we achieve critical coupling when the absorption rate matches the cavity decay rate ($\gamma_a = \gamma_d$). So for any given cavity there is always a critical absorptivity of its elements that will promote perfect absorption. It is interesting to note that deviating in any way from this condition reduces the overall absorption. That is, we have the (maybe) counter-intuitive result that for a cavity with a small decay rate (i.e. high quality factor $Q$) we may need to reduce the absorptivity of the cavity elements in order to increase the overall absorption. Graphene is a particularly promising material towards this direction as its internal absorptivity can be easily controlled and fine-tuned by electrostatic doping.

We explore the graphene tunability towards critical coupling in Fig. 1c-d, where we show the total and constituent absorptions as a function of graphene doping. We distinguish two cases, where the Au mirror is in direct contact with a Si layer (Fig. 1c) or with a SiO$_2$ layer (Fig. 1d), as shown by the inset schematics. We note that because the resonant wavelength shifts as we dope graphene, and to facilitate our discussion on critical coupling, for each different doping we scan the wavelength range and plot the point of maximum total absorption. We find that the simple change of Au mirror contact (Si or SiO$_2$) imposes dramatic changes between the two absorption graphs. Specifically, in case of direct contact with Si, the Au mirror exhibits higher absorption compared to the other case. Simply, at $\lambda = 1550$ nm the Au refractive index is $n_{Au} = 0.51 + i10.7$ and its single-pass absorption under normal incidence is $A_{Au} = 1 - R_\infty$, where the semi-infinite reflectance is $R_\infty = |n_{diel} - n_{Au}|^2/|n_{diel} + n_{Au}|^2$. Thus, the absorptivity in the Si/Au interface is 5.4%, about 2.2 times larger compared to the 2.5% of the SiO$_2$/Au interface. Consequently, the required SLG absorptivity for critical coupling should be about 2.2 times lower in the Si-terminated case, as is indeed shown in Fig. 1c-d. The lower Au absorptivity in SiO$_2$/Au implies larger room for tunability by graphene. Also, working in reflection mode the largest contrast is obtained in the SiO$_2$/Au case between critically-coupled graphene and highly doped graphene. Thus, the optimal configuration is with the smallest index dielectric in contact with the Au mirror.

For evaluating device performance we define the modulation depth either as MD[%] = $100(R_{max} - R_{min})/R_{max}$ or MD[dB] = $10\log_{10}(R_{max}/R_{min})$ and insertion loss either as IL[%] = $100(1 - R_{max})$ or IL[dB] = $-10\log_{10}(R_{max})$. Maximum reflectance $R_{max}$ is obtained at the highest graphene doping (e.g. at $E_F = 0.7$ eV) and minimum reflectance $R_{min}$ at critical coupling. The later varies depending on structure and graphene quality. In Fig. 2a we plot the modulation depth as a function of detuning $\delta E_F = E_F - E_F^c$ from the critical coupling condition, i.e. the detuning of $R_{min}$. Extremely large values can be obtained provided we can tune the graphene Fermi level with enough accuracy. Given the natural limitations to this, we explore the more realistic condition of having a tolerance in the applied gate voltage. In particular, the $E_F = 0.428$ eV doping needed for critical coupling (see Fig. 1d) results from $V_g = d_{SiO_2} e E_F^2 / (\varepsilon_0 \varepsilon_{SiO_2} \pi \hbar^2 v_F^2) = 165$ V, where $v_F = 10^6$ m/s the Fermi velocity, $d_{SiO_2} = 265$ nm the SiO$_2$ gate thickness and $\varepsilon_{SiO_2} = 3.9$ the gate dielectric constant. A tolerance of $\mathcal{D}E_F = 1$ meV (dotted lines in Fig. 2a) corresponds to $\mathcal{D}V_g = 0.8$ V, or a ~0.5% variation, while a $\mathcal{D}E_F = 10$ meV tolerance (full width of Fig. 1a) corresponds to ~5% of $V_g$. Our minimum reflectance is then $\langle R_{min}\rangle = (\mathcal{D}E_F)^{-1} \int R(\epsilon) d\epsilon$, where the integration is performed within $\mathcal{D}E_F$ around $E_F^c$. For our high quality graphene with $\tau = 200$ fs we find the modulation depth $\langle MD\rangle = 6.5 \times 10^4$ and $6.7 \times 10^2$ for 0.5% and 5% gating tolerance respectively.

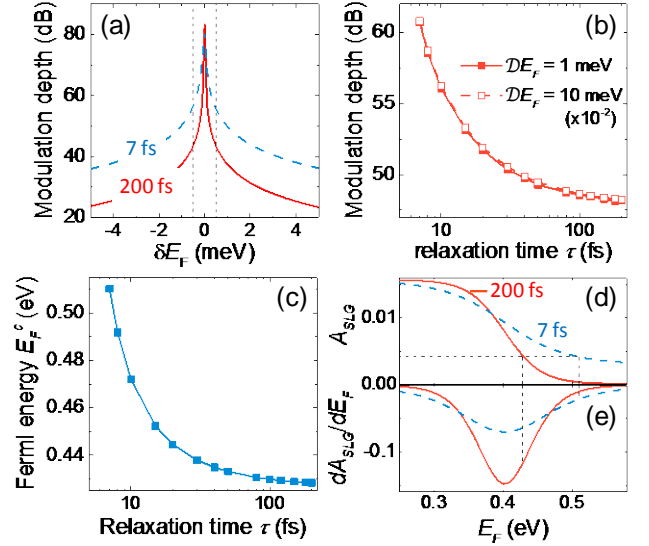

**Figure 2.** (a) Modulation depth MD[dB] for the device shown in Fig. 1a as a function of detuning $\delta E_F = E_F - E_F^c$ from $R_{min}$. (b) $\langle MD\rangle$ as a function of electron relaxation time assuming two tolerances on electrostatic doping. (c) Critical Fermi level $E_F^c$ for achieving critical coupling. (d) Graphene absorptivity for $\tau = 200$ fs (solid red) and $\tau = 7$ fs (dashed blue) (e) Absorption change with Fermi level for the two relaxation times above.

We next repeat the calculation for smaller electron relaxation times $\tau$ (corresponding to lower graphene mobilities according to[25] $\mu = e\tau v_F^2/E_F$). As an example, Fig. 2a also plots the modulation depth for $\tau = 7$ fs (i.e. $\mu \cong 140$ cm$^2$V$^{-1}$s$^{-1}$) found at critical doping $E_F^c = 0.51$ eV (for comparison, for $\tau = 200$ fs and critical doping $E_F^c = 0.428$ eV we find $\mu \cong 4650$ cm$^2$V$^{-1}$s$^{-1}$). While the peak value remains the same, there is a wider response curve, resulting in a much larger averaged modulation depth $\langle MD\rangle$. This is actually quite a consistent trend as shown in Fig. 2b where we plot $\langle MD\rangle$ as a function of $\tau$ for both 0.5% and 5% tolerances. More than an order of magnitude increase in modulation depth is obtained when we use lower quality graphene.

This is a remarkable result that significantly loosens graphene quality and fabrication requirements. To understand its origins, we plot in Fig. 2c the Fermi level of graphene required to reach critical coupling as a function of $\tau$. We note that a higher doping is required for lower $\tau$. In Fig. 2d we plot the absorptivity of graphene suspended within SiO$_2$ at $\lambda = 1550$ nm as a function of Fermi level, given by $A_{SLG} = 2\pi d_{gr} \varepsilon_2 / n_{SiO_2} \lambda$, where $d_{gr} = 0.335$ nm is the effective graphene thickness and $\varepsilon_2(\omega, E_F)$ the imaginary part of the graphene's dielectric function[22]. In order to keep the graphene absorptivity constant as we reduce $\tau$, we indeed need to increase the doping. What determines then the magnitude of ⟨MD⟩ is the derivative of absorptivity with doping, i.e. $dA_{SLG}/dE_F$, which is plotted in Fig. 2e. The two vertical lines mark $\tau = 200$ fs and $\tau = 7$ fs, showing about an order of magnitude decrease in slope for the latter case. However, the above result is not generic but rather specific to the requirements of the studied device. As seen in Fig. 2e, there are doping levels where $dA_{SLG}/dE_F$ is smaller for the high quality graphene. If critical coupling had occurred at these doping levels (e.g. if we had started with another device layout) then the high quality graphene would appear as the champion.

To explore the effect of device layout we vary the number of Bragg periods. In Fig.3 three different configurations with 2, 3 and 4 periods of Si/SiO$_2$ bilayers are evaluated (marked as R2, R3 and R4; in the R2 configuration because of the much higher $\gamma_d$ we needed high internal absorptivity and thus used the Si/Au configuration as in Fig. 1c). Alongside, we show the performance of a similar modulator (a symmetric one without the Au back mirror) operating in transmission mode (with 4, 5 and 6 Bragg periods, marked T4, T5 and T6). In this case the performance deteriorates quickly with reduced graphene mobility. This behavior is expected, since such a device requires graphene to be fully transparent in the *on* state for maximum response. This is in contrast to the proposed modulator in reflection mode which shows enhanced performance for lower quality graphene. We should note that in Fig3a we present the results using a 0.5% tolerance on gate voltage (curves should be lowered by 20 dB for a 5% tolerance).

Adding Si/SiO$_2$ bilayers on both sides improves the modulation depth in transmission mode because of a lower $T_{\min}$ in the *off* state. This however carries a major drawback as it also leads to an increase in the device's insertion loss as seen in Fig3b. In reflection mode, in contrast, adding Si/SiO$_2$ bilayers reduces the modulation depth (because of a narrower $R_{\min}$, see Fig. 2a) but decreases the insertion loss. A 3-(Si/SiO$_2$) bilayer system in reflection mode overall outperforms all the other examined devices in all the desired figure of merits, even with a 5% tolerance in gate voltage, and most importantly, without any restrictions on graphene quality.

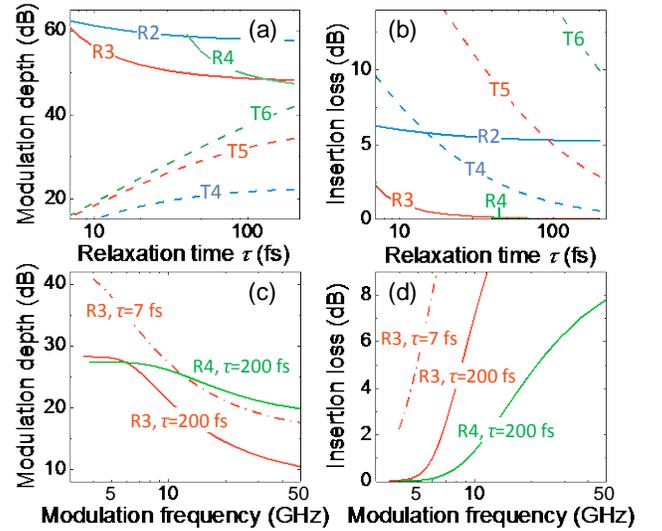

**Figure 3.** (a) ⟨MD⟩ and (b) IL for 3 reflective devices consisting of 2, 3 and 4 Bragg periods on each side (R2, R3 and R4) and 3 transmitting devices consisting of 4, 5 and 6 Bragg periods on each side (T4, T5 and T6). (c) ⟨MD⟩ *vs* MF and (d) IL *vs* MF trade-offs for the R3 and R4 devices.

The optics-limited operating frequency is given by[40] $f_{\text{opt}} = c/(n_{\text{SiO}_2} 2LQ)$, where $c/n_{\text{SiO}_2}$ is the speed of light in the dielectric, $2L = 1060$ nm the roundtrip length of the cavity (the middle 2 SiO$_2$ layers in our case) and $Q$ the cavity quality factor defined as $Q = \omega_0/\Delta\omega$, where $\omega_0$ is the cavity mode frequency and $\Delta\omega$ the FWHM. For our case of 3 bilayers we find $Q{\sim}325$ and $f_{\text{opt}}{\sim}600$ GHz. This is close to the fundamental speed limits posed by the photocarrier generation and relaxation processes which are up to picosecond timescales[8]. The electronics-limited operating frequency is however much smaller, defined by $f_{\text{el}} = (2\pi RC)^{-1}$, where $R_{\text{el}}$ is the system's resistance and $C = \varepsilon_0 \varepsilon_{SiO_2} A/d_{SiO_2}$ its capacitance. For an $A = 50 \times 50$ μm$^2$ square area device (covering a typical optical beam size) we estimate $C = 0.33$ pF and $R_{\text{el}}{\sim}0.1$ kΩ and $R_{\text{el}}{\sim}2.4$ kΩ for the $\tau = 200$ fs and $\tau = 7$ fs cases respectively, where $R_{\text{el}} = \sigma^{-1} = ne\mu$ with $n = E_F^2/\pi\hbar^2 v_F^2$ the charge density and $\mu$ as defined above. From these we find $f_{\text{el}}{\sim}5$ GHz and ${\sim}0.2$ GHz. Apparently, higher quality graphene and/or smaller device footprint lead to higher modulation frequencies.

Given the exceptionally high MD and low IL predicted in these structures, there is some room to further improve MF by trading with MD and IL. In particular, for charging between gate voltages $V_1$ and $V_2$ the instantaneous potential is $V = V_2 + (V_1 - V_2)e^{-t/RC}$. Assuming the charging (or discharging) to be complete when it reaches within $\mathcal{D}V_g/2$ from its target value, we estimate the modulation frequency as $f_{el} = [2RC\ln(2\Delta V/\mathcal{D}V_g)]^{-1}$, where $\Delta V = V_2 - V_1$. A reduced $\Delta V$ corresponds to a higher $f_{el}$ but also to a reduced $\Delta E_F$ and thus to smaller $\langle MD \rangle$ and larger IL. These trade-offs for the R3 and R4 devices are shown in Fig. 3c-d using $\mathcal{D}V_g = 5\%$. A great design flexibility is found, in particular for higher quality graphene inside a higher Bragg-count device.

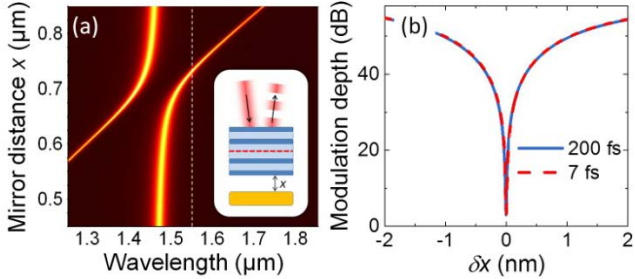

**Figure 4.** (a) Overall absorption *vs* mirror separation $x$ and wavelength. The inset schematic shows the device configuration. Color scale is from 0-1. (b) $\langle MD \rangle$ as a function of detuning from critical separation $x_0$.

We have shown that the smallest deviation from critical coupling has a major consequence in the device reflection. This fact can be exploited to design a sensitive acousto–optic modulator. We separate the Au mirror from the rest of the device by a distance $x$, creating a secondary cavity as seen in the inset of Fig 4a. For every distance $x$ we tune the graphene Fermi level into its corresponding critical coupling value $E_F^c$ and plot in Fig. 4a the total absorption as a function of mirror separation $x$ and wavelength. A strong anti-crossing behavior is found at around $x \sim 700$ nm, due to interference between the $SiO_2$ cavity and the mirror cavity, making up for an extremely sensitive response. We pick $x_0 = 735$ nm, where perfect absorption is observed for $\lambda = 1550$ nm and plot in Fig. 4b the $\langle MD \rangle$ (assuming $\mathcal{D}V_g = 0.5\%$; gating is fixed in this case and thus easily tuned to high accuracy) as a function of mirror deflection $\delta x = x - x_0$. Strong reflections with increasing $\delta x$ make up for very high $\langle MD \rangle$ values. For example, a tiny $\delta x = 1$ Å results in a massive increase in reflectance $R = 10^3 \langle R_0 \rangle$ corresponding to a ~30 dB/Å sensitivity (only drawback is the small overall reflection and thus large IL). By the same token, a small change in the refractive index inside the mirror cavity will also produce a measurable change in reflectance. Since the optical length is $nx$, a $\delta n/n = \delta x/x$ will produce the same $\delta R/R$. Using the numbers of Fig. 4b we estimate an index sensitivity of the order of $10^7$ %/RIU. We note that these high sensitivities are independent of graphene mobility, as seen in Fig. 4b, because the only requirement is to find the appropriate $E_F^c$ to tune into critical coupling.

In conclusion, we have shown that critical coupling is a versatile tool in achieving deep and fast optical modulation: a graphene-loaded Bragg resonator device operating in reflection mode can always be tuned into critical coupling irrespective of graphene quality. The key to this is our ability to fine tune the graphene electrostatic doping in order to match the cavity decay and absorption rates, without having to alter the device architecture. This scheme consistently yields large modulation depth (~100%) at small insertion loss (<1 dB) with high modulation frequency (>1 GHz), even when graphene electron relaxation time falls below 10 fs. This versatile mechanism can in addition be extended to include other operation formats, such as a sensitive acousto-absorption modulator (~30 dB/Å) or a sensitive index sensor ($10^7$ %/RIU). All the considerations and designs studied here for $\lambda = 1550$ nm can also be easily extended throughout the midIR spectrum by appropriately choosing the materials and scaling their thicknesses.

This project has received funding from the European Union's Horizon 2020 research and innovation programme under grant agreement No 696656.

[1] G. P. Agrawal, *Fiber-optic communication systems* (John Wiley & Sons, Hoboken, NJ, 2010).
[2] G. T. Reed and A. P. Knights, *Silicon photonics: an introduction* (John Wiley & Sons, Chichester, UK, 2004).
[3] G. T. Reed, G. Mashanovich, F. Y. Gardes, and D. J. Thomson, Nat. Photonics 4, 518 (2010).
[4] A. K. Majumdar and J. C. Ricklin (ed.), *Free-space laser communications* (Springer, New York, NY, 2008).
[5] H. Kaushal and G. Kaddoum, IEEE Commun. Surveys & Tutorials **19**, 57 (2017).
[6] A.C. Ferrari et al., Nanoscale 7, 4598 (2015).
[7] A. H. Castro Neto, F. Guinea, N. M. R. Peres, K.S. Novoselov, and A. Geim, Pev. Mod. Phys. 81, 109 (2009).

[8] A. Urich, K. Unterrainer, and T. Mueller, Nano Lett. 11, 2804 (2011).
[9] F. Bonaccorso, Z. Sun, T. Hasan, A.C. Ferrari, Nat. Photon. 4, 611 (2010).
[10] F. Koppens, T. Mueller, P. Avouris, A.C. Ferrari, M. Vitiello, M. Polini, Nat. Nanotech. 9, 780 (2014).
[11] M. Liu, X. Yin, E. Ulin-Avila, B. Geng, T. Zentgraf, L. Ju, F. Wang, and X. Zhang, Nature 474, 64 (2011).
[12] C.T. Phare, Y.-H. D. Lee, J. Cardenas, and M. Lipson, Nat. Photonics 9, 511 (2015)
[13] N. Youngblood, Y. Anugrah, R. Ma, S. J. Koester, and M. Li, Nano Lett. 14, 2741 (2014)
[14] M. Mohsin, D.l Schall, M.Otto, A. Noculak, D. Neumaier, and H. Kurz, Opt. Express 22, 15292 (2014)
[15] D. Ansell, I.P. Radko, Z. Han, F.J. Rodriguez, S.I. Bozhevolnyi, and A.N. Grigorenko, Nat. Commun. 6, 8846 (2015).
[16] C.-C. Lee, S. Suzuki, W. Xie, and T. R. Schibli, Opt. Express 20, 5264 (2012).
[17] F. J. Rodriguez, D. E. Aznakayeva, O. P. Marshall, V. G. Kravets, and A. N. Grigorenko, Adv. Mater. 29, 1606372 (2017).
[18] T. Sun, J. Kim, J. M. Yuk, A. Zettl, F. Wang, and C. Chang-Hasnain, Opt. Express 24, 26035 (2016).
[19] A. Majumdar, J. Kim, J. Vuckovic, and F. Wang, Nano Lett. 13, 515 (2013).
[20] R.R. Nair, P. Blake, A.N. Grigorenko, K.S. Novoselov, T.J. Booth, T. Stauber, N.M.R. Peres, and A.K. Geim, Science 320, 1308 (2008).
[21] K. F. Mak, M. Y. Sfeir, Y. Wu, C. H. Lui, J.A. Misewich, and T. F. Heinz, Phys. Rev. Lett 101, 196405 (2008).
[22] T.J. Echtermeyer, S. Milana, U. Sassi, A. Eiden, M. Wu, E. Lidorikis, and A.C. Ferrari, Nano Lett. 16, 8 (2016).
[23] F. Wang, Y. Zhang, C. Tian, C. Girit, A.Zettl, M.Crommie, and Y. R. Shen, Science 320, 206 (2008).
[24] Z. Q. Li, E. A. Henriksen, Z. Jiang, Z. Hao, M. C. Martin P. Kim, H. L. Stormer, and D. N. Basov, Nat. Physics 4, 532 (2008).
[25] M. Jablan, H. Buljan, and M. Soljačić, Phys. Rev. B 80, 245435 (2009).
[26] A.N. Grigorenko, M. Polini, and K.S. Novoselov, Nature Photonics 6, 749 (2012).
[27] F. J. García de Abajo, ACS Photon. 1, 135 (2014)
[28] S. Thongrattanasiri, F. H. L. Koppens, and F. J. García de Abajo, Phys. Rev. Lett. 108, 047401 (2012).
[29] A. Ferreira, N. M. R. Peres, R. M. Ribeiro, and T. Stauber, Phys. Rev. B 85, 115438 (2012).
[30] M. Furchi, A. Urich, A. Pospischil, G. Lilley, K. Unterrainer, H. Detz, P. Klang, A. M. Andrews, W. Schrenk, G. Strasser, T. Mueller, Nano Lett. 12, 2773 (2012).
[31] M. Engel, M. Steiner, A. Lombardo, A. C. Ferrari, H. v. Löhneysen, P. Avouris, and R. Krupke, Nat. Commun. 3, 906 (2012).
[32] W.Wang, A. Klots, Y. Yang, W. Li, I. I. Kravchenko, D. P. Briggs, K. I. Bolotin, and J. Valentine, Appl. Phys. Lett, 106, 181104 (2015).
[33] R. Yu, V. Pruneri, and F.J. García de Abajo, Scientific Reports 6, 32144 (2016).
[34] T. J. Echtermeyer, L. Britnell, P. K. Jasnos, A. Lombardo, R. V. Gorbachev, A. N. Grigorenko, A. K. Geim, A. C. Ferrari, K. S. Novoselov, Nat. Commun. 2, 458 (2011).
[35] V.G. Kravets, F. Schedin, R. Jalil, L. Britnell, K.S. Novoselov, and A.N. Grigorenko, J. Phys. Chem. C, 116, 3882 (2012).
[36] B.D. Thackray, P.A. Thomas, G.H. Auton, F.J. Rodriguez, O.P. Marshall, V.G. Kravets, and A.N. Grigorenko, Nano Lett. 15, 3519 (2015).
[37] F. Schedin, E. Lidorikis, A. Lombardo, V.G. Kravets, A.K. Geim, A.N. Grigorenko, K.S. Novoselov, and A.C. Ferrari, ACS Nano 4, 5617 (2010)
[38] S. Fenghua, C. Yihang, H. Peng, and P. Tassin, Small 11, 6044 (2015).
[39] B. Vasić, and R. Gajić, Opt. Lett. 39, 6253 (2014).
[40] R. Yu, V. Pruneri, and F.J. García de Abajo, ACS Photonics 2, 550 (2015).
[41] A. Yariv, IEEE Photonic Technol. Lett. 14, 483 (2002).
[42] J.R. Tischler, M.S. Bradley, and V. Bulovic, Opt. Lett. 31, 2045 (2006)
[43] J.R. Piper, V. Liu, and S. Fan, Appl. Phys. Lett. 104, 251110 (2014).
[44] J.R. Piper and S. Fan, ACS Photonics, 1, 347 (2014).
[45] G. W. Hanson, J. Appl. Phys. 104, 084314 (2008).
[46] H. A. Haus, Waves and Fields in Optoelectronics (Prentice-Hall, Englewood Cliffs, NJ, 1984).